\documentstyle[prc,aps,subeqnarray]{revtex}
\newcommand \be{\begin{eqnarray}}
\newcommand \ee{\end{eqnarray}}

\begin{document}

\title{ Restoration of broken symmetries in Self-Consistent RPA }

\author{A. Rabhi$^{1,2}$, P. Schuck$^{3}$, R. Bennaceur$^{1}$, G. Chanfray$^{2}$, J. Dukelsky$^{4}$ \\
{\small $^1$Laboratoire de Physique de la Mati\`ere Condens\'ee}, \\
{\small Facult\'e des Sciences de Tunis, Campus Universitaire, Le Belv\'ed\`ere-1060, Tunisia} \\
{\small $^2$IPN-Lyon, 43Bd du 11 novembre 1918, F-69622 Villeurbanne Cedex, France}\\
{\small $^3$Institut de Physique Nucl\'eaire, IN2P3-CNRS, Universit\'e Paris-Sud, F-91406 Orsay Cedex, France}\\
{\small $^4$Instituto de Estructura de la Materia, C.S.I.C., Madrid, Spain}}
\date{\today}
\maketitle

\begin{abstract}
It is shown that the Self-Consistent RPA (SCRPA) approach allows in a very natural way to restore
symmetries, spontaneously broken on the mean field level. This is achieved via the introduction of
a second Lagrange multiplier which constrains the variance of the symmetry operator to zero. This
important feature of SCRPA, here pointed out for the first time, is illustrated employing a simplified
model of the nuclear superfluidity.  \\
\end{abstract}


Extensions of RPA theory, based on the Equation of Motion (EOM) approach have by now a quite long
history. They, to a great deal, have been developed in nuclear physics. It started out with the
work of Hara who included ground state correlations in the Fermion occupation numbers \cite{1}.
More systematic was the consequent work by Rowe and coworkers (see the review by D. J. Rowe \cite{2}).
The same theory was developed using the Green's Function method by one of the present authors in
\cite{3}. Independently the method was proposed by R. Zimmermann and G. R\"opke plus coworkers
using a graphical construction \cite{4}. These authors called their method Cluster-Hartree-Fock
(CHF) but it is equivalent to Self-Consistent RPA (SCRPA). The latter approach has recently been
further developed by Dukelsky and Schuck in a series of papers \cite{5,6,7,8}.
However, also other authors
contributed actively to the subject \cite{9}. A number of remarkable results have been obtained with
SCRPA in non trivial models where comparison with exact solutions was possible \cite{8,10}.
For instance for the exactly solvable many level pairing model of Richardson \cite{11} SCRPA provides
very accurate results for the ground state and the low lying part of the spectrum \cite{8}. \\
One important problem which remained little explored so far is the question how to deal with situations
when there appears a continuously broken symmetry on the mean field level like nuclear deformation
or superfluidity. We know that the standard RPA restores the broken symmetry in conjunction with a
self-consistent mean-field single particle basis \cite{12}. Complying with the Goldstone theorem is very
important, since this assures simultaneously that conservation laws and sum rules are fulfilled.
Unfortunately, it is a notoriously difficult problem that these properties are not easy to maintain
when going beyond the standard RPA approach. SCRPA can be derived variationally and it couples in a
non linear way the ordinary mean field to the quantal fluctuations \cite{5,6}.
If a symmetry is spontaneously broken it is customary in the Hartree-Fock (HF) or 
Hartree-Fock-Bogoliubov (HFB) approaches to fix at least the mean value of the symmetry operator with a
Lagrange multiplier. There have been some attempts, with not very convincing outcome,
to fix also its second moment (see for example the method proposed by Lipkin and Nogami in the
case of pairing \cite{13}). HF and HFB approaches lead to one body mean field equations and it does
not seem very natural to fix at the same time the mean value of a one body and a two body operator,
that is e.g. the second moment of the symmetry operator. On the other hand in SCRPA which couples
one body and two body mean field equations (see below) the requirement of simultaneously imposing
the correct first and second moment is quite natural. This idea has to our knowledge not been explored
so far and we present it here for the first time for a very simple model case of nuclear superfluidity.
We will develope the general theory in a future work but in this short letter we think it appropriate
to work out the theory in the very transparent case of the seniority model \cite{12} where we indeed
will show that SCQRPA, as standard QRPA, completely restores the symmetry. However, contrary to
standard QRPA, in the case of SCQRPA, because it is variational, this is equivalent to solving the model
exactly.\\    
The Hamiltonian of the seniority model is given by :
\begin{equation}
H=- G\Omega S^{\dagger}S
\label{1}
\end{equation}
where G is the pairing strength, $2\Omega$ is the degeneracy of the level and the quasi-spin
raising and lowering operators are given by :
\begin{equation}
S^{\dagger}=\frac{1}{\sqrt{\Omega}}\sum_{m>0}a^{\dagger}_{m}a^{\dagger}_{-m}, \quad S=(S^{\dagger})^{\dagger}
\label{2}
\end{equation}
with m the quantum numbers of the magnetic substates. Since we will work with the quasiparticle RPA
(QRPA), we will use instead of (\ref{1}) the following constrained Hamiltonian:
\begin{equation}
H^{\prime}=H-\mu_{1}\hat{N}-\mu_{2}{\hat{N}}^{2}
\label{3}
\end{equation}
where $\hat{N}=\sum_{m>0}(a^{\dagger}_{m}a_{m}+a^{\dagger}_{-m}a_{-m})$ is the particle number operator,
$\mu_{1}$ and $\mu_{2}$ are two Lagrange multipliers allowing to fix the mean values
$N=\langle \hat{N} \rangle $ and $ N^{2}=\langle \hat{N}^{2}\rangle $ as we indicated above.
We will transform $H^{\prime}$ to usual Bogoliubov quasiparticles \cite{12} :
\begin{equation}
a^{\dagger}_{m}=u\alpha^{\dagger}_{m}+v\alpha_{-m}, \quad u^{2}+v^{2}=1 
\label{4}
\end{equation}
and define new quasi-spin operators as :
\begin{equation}
P^{\dagger}=\frac{1}{\sqrt{\Omega}}\sum_{m>0}\alpha^{\dagger}_{m}\alpha^{\dagger}_{-m}
, \quad P=(P^{\dagger})^{\dagger}
\label{5}
\end{equation}
where the quasiparticle number operator is given by :
\begin{equation}
\hat{n}_{q}=\sum_{m>0}(\alpha^{\dagger}_{m}\alpha_{m}+\alpha^{\dagger}_{-m}\alpha_{-m}).
\label{6}
\end{equation}
We will not repeat here how the Hamiltonian $H^{\prime}$ is expressed in the quasiparticle operators,
since this is given in the text books \cite{12}. However, we briefly want to outline the main steps
of the SCQRPA approach. The RPA excited states are, as usual, obtained as :
\begin{equation}
|\nu \rangle=Q^{\dagger}|RPA\rangle
\label{7}
\end{equation}
where $|RPA\rangle $ is the correlated RPA ground state defined via the vacuum condition :
\begin{equation}
Q|RPA\rangle=0,
\label{8}
\end{equation} 
and the QRPA excitation operator is given by \cite{6} :
\begin{equation}
Q^{\dagger}=xP^{\dagger}-yP, \quad x^{2}-y^{2}=1.
\label{9}
\end{equation}
It is directly checked that relation (\ref{9}) can be inverted, that is :
\begin{equation}
P^{\dagger}=xQ^{\dagger}+yQ
\label{10}
\end{equation}
and therefore with condition (\ref{8}) a certain number of expectation values are readily expressed by
the RPA amplitudes $x$ and $y$ :
\be
\langle P^{\dagger}P \rangle &=&y^{2}(1-\frac{\langle \hat{n}_{q} \rangle}{\Omega}), \cr
\langle PP^{\dagger} \rangle&=&x^{2}(1-\frac{\langle \hat{n}_{q} \rangle}{\Omega}), \cr
\langle P^{\dagger}P^{\dagger} \rangle &=& \langle PP \rangle = xy(1-\frac{\langle \hat{n}_{q}
\rangle}{\Omega}) \label{11}
\ee
where $\langle ... \rangle = \langle RPA|...|RPA\rangle/\langle RPA|RPA\rangle$.
With (\ref{7}) and (\ref{8}) it is standard procedure using e.g. the EOM approach to derive the RPA
secular equation \cite{2} :
\begin{equation}
\langle [\delta Q,[H^{\prime},Q^{\dagger}]]\rangle=E\langle [\delta Q,Q^{\dagger}]\rangle
\label{12}
\end{equation} 
or explicitly :
\begin{equation}
\pmatrix{ A & B \cr -B & -A \cr }
\pmatrix{ x \cr  y \cr } = E \pmatrix{ x \cr y \cr };\quad x^{2}-y^{2}=1
\label{13}
\end{equation}
where,
\be
A&=&\frac{\langle [P,[H^{\prime},P^{\dagger}]]\rangle }{1-\frac{\langle \hat{n}_{q} \rangle}{\Omega}}\cr
&=&2(G-4\mu_{2})\bigg \{(2v^{2}(1-v^{2})-1)xy-v^{2}(1-v^{2})(1+2y^{2}-\Omega\frac{1-2\frac{\langle
\hat{n}_{q}\rangle}{\Omega}+\frac{\langle\hat{n}_{q}^{2}\rangle}{\Omega^{2}}}{1-\frac{\langle 
\hat{n}_{q} \rangle}{\Omega}})\bigg \}, \label{a14}\\
B&=&\frac{\langle [P^{\dagger},[H^{\prime},P^{\dagger}]]\rangle}{1-\frac{\langle \hat{n}_{q}
\rangle}{\Omega}} \cr
&=&2(G-4\mu_{2})\bigg\{(6v^{2}(v^{2}-1)+1)xy-v^{2}(1-v^{2})(1+2y^{2}-\Omega\frac{1-2\frac{\langle
\hat{n}_{q}\rangle}{\Omega}+\frac{\langle\hat{n}_{q}^{2}\rangle}{\Omega^{2}}}{1-\frac{\langle 
\hat{n}_{q} \rangle}{\Omega}})\bigg\}.
\label{a15}
\ee
It is important to note that eqs (\ref{12}-\ref{a15}) can also be derived from a variational principle
using the following functional \cite{14}, which is a special combination of the energy weighted sum
rule \cite{12} :  
\be
{\cal E }[x,y]=\frac{\sum_{\nu}(E_{\nu}-E_{0})|\langle \nu|Q^{\dagger}|0\rangle|^{2}-
\sum_{\nu'}(E_{0}-E_{\nu'})|\langle \nu'|Q|0\rangle|^{2}}{\sum_{\nu}|\langle
\nu|Q^{\dagger}|0\rangle|^{2}-\sum_{\nu'}|\langle \nu'|Q|0\rangle|^{2}}
\label{16}
\ee
with $E_{\nu}$ and $|\nu\rangle$ eigenvalues and eigenstates of $H^{\prime}$. \\
  
Since (\ref{9}) can be considered as a Bogoliubov transformation of Fermion pair operators we can
interpret the SCQRPA equations as the corresponding mean field equations. In other words the SCQRPA equations are
the mean field equations for a gas of quantal pair fluctuations. The standard mean field equations of the
one body type are, as usual, derived in minimizing the constrained ground state energy
$\langle RPA|H^{\prime}|RPA\rangle\equiv E^{\prime}_{0}$ with respect to the mean field amplitudes
$u, v$ :
\begin{equation}
\frac{\partial E^{\prime}_{0}}{ \partial v} \equiv \langle [H^{\prime},P]\rangle =0.
\label{17}
\end{equation}
This last equation (which can be verified in evaluating l.h.s and r.h.s separately) is very interesting,
since it is equivalent to
$\langle [H',Q]\rangle=i\frac{\partial}{\partial t}\langle Q(t)\rangle=0$ which should be fulfilled
at equilibrium and consequently constitutes a further relation of the EOM approach, first put forward
in \cite{5}. With the normalization of the amplitudes $u,v$ (see (\ref{4})) one can readily rewrite
(\ref{17}) in the more common form :
\begin{equation}
\pmatrix{ h & \Delta \cr \Delta & -h \cr }
\pmatrix{ u \cr  v \cr } = \epsilon \pmatrix{ u \cr v \cr }
\label{18}
\end{equation}
where,
\begin{subeqnarray}
h&=&-(G+4\mu_{2}\Omega)v^{2}-\mu_{1},  \label{19}\slabel{h}\\
\Delta&=&(G\Omega+4\mu_{2})uv-(G-4\mu_{2})uv \bigg\{ 2(xy+y^{2})+\frac{\langle \hat{n}_{q}\rangle
-\frac{\langle \hat{n}_{q}^{2}\rangle}{\Omega}}{1-\frac{\langle
\hat{n}_{q}\rangle}{\Omega}}\bigg\},  \slabel{Delta}\\
\epsilon&=&\sqrt{h^{2}+\Delta^{2}} \slabel{epsilon}.
\end{subeqnarray}
We see that once we replace in (\ref{19}) $| RPA \rangle $ by the uncorrelated HFB vacuum with 
$$ y = \langle \hat{n}_{q} \rangle = \langle \hat{n}_{q}^{2} \rangle =0\quad \textrm{and}\quad \mu_{2}=0 $$
we obtain from (\ref{18}) the standard HFB equations. However, in keeping with the correlated vacuum
the Fermionic mean field equations (\ref{18}) get coupled to the mean field equations (\ref{13}) for the
quantal pair fluctuations which is a very gratifying aspect of SCRPA theory.\\
In addition to the two mean field equations (\ref{13}) and (\ref{18}) we have two further equations
which, in principle, allow us to find the Lagrange multipliers $\mu_{1}$, $\mu_{2}$ (see, however, below) :
\begin{subeqnarray}
N&=&\langle \hat{N}\rangle =(u^{2}-v^{2})\langle\hat{n}_{q}\rangle+2\Omega v^{2},
\label{20}\slabel{seq1}\\
N^{2}&=&\langle \hat{N}^{2}\rangle =(u^{2}-v^{2})^{2}\langle\hat{n}_{q}^{2}\rangle 
+8\Omega u^{2}v^{2}(1-\frac{\langle\hat{n}_{q}\rangle}
{\Omega})(xy+y^{2})+4\Omega v^{2}(u^{2}+\Omega v^{2}) \cr
&+&4v^{2}(\Omega(u^{2}-v^{2})-u^{2})\langle\hat{n}_{q}\rangle \slabel{seq2}
\end{subeqnarray}
We again see that eqs (\ref{20}) reduce to the standard expressions, once, as in the HFB approximation,
we pose $y=\langle\hat{n}_{q}\rangle=\langle\hat{n}_{q}^{2}\rangle=0$. In the particular case of the
seniority model the number equation (\ref{seq1}) in the HFB approximation determines the amplitudes
$u,v$ and then no freedom is left to impose $\Delta N=\sqrt{\langle \hat{N}^{2} \rangle -{\langle \hat{N}
\rangle}^{2}}=0$. However, in the present more general approach
there is more freedom and we will be able to satisfy the relation $\Delta N= 0$. Before getting
to this point, we first have to show how the so far open quantities $\langle \hat{n}_{q}\rangle $ and
$\langle\hat{n}_{q}^{2}\rangle$ can be expressed via our four unknowns $v$, $y$ , $\mu_{1}$ and
$\mu_{2}$ (the two other amplitudes $u$ and $x$ being determined via the corresponding normalization
conditions in (\ref{4}) and (\ref{9})). The determination of those two quantities is one of the
difficulties in the SC(Q)RPA approach \cite{5,6}. However, this problem has found an elegant solution
in the early work of \cite{15}.
There is no place here to go into the details of the derivation. Let us simply state that applied
to the present model one is able to expand $\hat{n}_{q}=\sum_{i=0}c_{i}(P^{\dagger})^{i}(P)^{i}$ and also
$\hat{n}_{q}^{2}=\sum_{i=0}b_{i}(P^{\dagger})^{i}(P)^{i}$ where the coefficients $c_{i}$,$b_{i}$ have
definite expressions. Explicitly this yields \cite{16} :
\begin{subeqnarray}
\hat{n}_{q}=\frac{2}{\Omega}P^{\dagger}P+\frac{2}{\Omega^{2}(\Omega-1)}{P^{\dagger}}^{2}P^{2}+
\frac{2}{\Omega^{3}(\Omega-1)(\Omega-2)}{P^{\dagger}}^{3}P^{3}+.... \label{21}\slabel{seq11}\\
\hat{n}_{q}^{2}=\frac{4}{\Omega}P^{\dagger}P+\frac{4(\Omega+1)}{\Omega^{2}(\Omega-1)}{P^{\dagger}}^{2}P^{2}+
\frac{8(\Omega+1)}{\Omega^{3}(\Omega-1)(\Omega-2)}{P^{\dagger}}^{3}P^{3}+.....\slabel{seq22}
\end{subeqnarray}
\linebreak[4]
We see that between (\ref{seq11}) and (\ref{seq22}) exists, at any step of the expansion, the following relation :
\begin{equation}
\frac{\Omega}{2}(P^{\dagger}P+PP^{\dagger})+(\frac{\hat{n}_{q}-\Omega}{2})^{2}=
\frac{\Omega}{2}(\frac{\Omega}{2}+1).
\label{30}
\end{equation}
It should be noted that eq (\ref{30}) is just the Casimir relation for the quasi-spins of the
seniority model. Using (\ref{10}), (\ref{8}) we obtain from (\ref{21}), (\ref{30}) two eqs for $\langle\hat{n}_{q}\rangle$ and
$\langle\hat{n}_{q}^{2}\rangle $ which can be solved in terms of $v$ and $y$ amplitudes and therefore the two mean field eqs (\ref{13})
and (\ref{18}) together with (\ref{seq1}) and (\ref{seq2}) represent four coupled nonlinear equations for
our unknowns $v$, $y$, $\mu_{1}$ and $\mu_{2}$. Usually the number equations (\ref{seq1}) and (\ref{seq2})
are to be used for the determination of the chemical potential $\mu_{1}$ and the second Lagrange
multiplier $\mu_{2}$ and the mean field equations for the amplitudes $u, v$ and $x, y$.
In the present case it is, however, more convenient to invert the role of mean field and number
equations, since eqs (\ref{20}) do not depend on the Lagrange multipliers and therefore readily allow
to determine $v^{2}$ and $y^{2}$ as a function of the particle number $N$.
Inversely the two mean field eqs (\ref{13}), (\ref{18}) are linear in $\mu_{1}$ and $\mu_{2}$ and for
instance it is seen that (\ref{13}) directly yields :
\begin{equation}
\mu_{2}=\frac{G}{4}  
\label{22}
\end{equation}
independent of the particle number $N$. Considering, the well known exact expression for the ground state
energy of the model \cite{12}:
\begin{equation}
E_{0}=-\frac{G}{2}(\Omega+1)N+\frac{G}{4}N^{2}
\label{23}
\end{equation}
we see from $\mu_{2}=\frac{\partial E_{0}}{\partial N^2}$ that (\ref{22}) gives the exact value for the
second Lagrange multiplier $\mu_{2}$. For the chemical potential $\mu_{1}$ we obtain from (\ref{18})
\begin{equation}
\mu_{1}=(G-4\mu_{2})\bigg\{(\Omega-1)v^{2}+(1-2v^{2})(xy+y^{2}+\frac{1}{2}\frac{\langle\hat{n}_{q}\rangle
-\frac{\langle\hat{n}_{q}^{2}\rangle}{\Omega}}{1-\frac{\langle
\hat{n}_{q}\rangle}{\Omega}})\bigg\}-\frac{G}{2}\Omega-2\mu_{2}.
\label{24}
\end{equation}
With relation (\ref{22}) this gives $\mu_{1}=-\frac{G}{2}(\Omega+1)$ which again is the exact value.
Furthermore, with (\ref{22}) we have from (\ref{a14}), (\ref{a15}) that $ A=B=0 $ and therefore the RPA
eigenvalue $E=0$.
This means that, as in standard QRPA, SCQRPA yields a Goldstone mode at zero energy which in nuclear
physics usually is called the spurious mode \cite{12}. This feature is very rewarding, since it signifies
that the particle number symmetry is exactly restored. \\
It is well known that restoration of good particle number implies in this very simple model case that
the model is solved exactly \cite{12}. We have already seen that one obtains the exact values for
$\mu_{1}$ and $\mu_{2}$.
We now will show that one also obtains the exact value for the ground state
energy (and therefore for the whole band of ground state energies). This goes as follows.
For the expectation value of $H$ of eq (\ref{1}) in the RPA ground state, using (\ref{8}), (\ref{10}) and the quasiparticle
representation for $H$, we can write 
\be
E_{0}&=&\langle H \rangle \cr
&=&-\frac{G}{2}(\Omega+1)\{(1-2v^{2})\langle \hat{n}_{q} \rangle +2\Omega v^{2}\}
+\frac{G}{4}\bigg\{(1-2v^{2})^{2}\langle \hat{n}_{q}^{2} \rangle \cr
&+&[4v^{2}(\Omega(1-2v^{2})-1+v^{2})-8(1-v^{2})v^{2}(xy+y^{2})]\langle \hat{n}_{q} \rangle \cr
&+&8\Omega(1-v^{2})v^{2}(xy+y^{2})+4\Omega v^{2}(1-v^{2}+\Omega v^{2})\bigg\}. 
\label{25}
\ee
In this expression we have used the relation 
$4y^{2}(\Omega- \langle \hat{n}_{q} \rangle)=2(\Omega+1) \langle \hat{n}_{q} \rangle
-\langle \hat{n}_{q}^{2} \rangle$ which follows from (\ref{30}).
Using the expression for $N$ and $ N^{2}$ of (\ref{seq1}) and (\ref{seq2}) once more, we see that the exact
expression (\ref{23}) is recovered. \\

Before concluding several remarks are in order. One may have noticed that the model is solved exactly
independent on how far we push the expansion (\ref{21}). This is, of course, particular to the simplicity
of the present model. As long as the theory is sufficiently general to fulfill the two number 
equations (\ref{seq1}) and (\ref{seq2}), the model will always be solved exactly in the present approach.
One also notes that we have developed our theory without ever using the RPA ground state wave-function
explicitly. Equation (\ref{8}) was used only implicitly. In the present case the vacuum condition can
be solved and an explicit expression for $|RPA\rangle$, quite analogous to the ones found in other
simplified models \cite{5}, can be given. We will not present this here, since it turned out in more realistic
cases that the explicit construction of the RPA ground state is extremely difficult, if not impossible.
We would like to point out, however, that also in HFB theory, in principle, the explicit knowledge of
the wave function is never necessary and for all relevant questions it is sufficient to deal with the
vacuum relation $\alpha |HFB\rangle =0$. \\

In conclusion we showed in this work how in the case of a symmetry broken situation the introduction of
a second constraint which yields the correct value (zero) for the variance of the symmetry operator 
allows to restore exactly the symmetry, and thus the conservation laws, in the SCRPA approach. This was
demonstrated using a simplified pairing model. It turns out that this solves the model exactly.
Far from being over simplified for the present purpose we think, on the contrary, that the model
perfectly illustrates the flexibility of the SCRPA to accommodate the additional constraint of putting
the variance of the symmetry operator to zero. Indeed in this model, precisely because of its
simplicity, there is within HFB no freedom for a constraint other than the single one for the mean value of the
particle number. We think that the possibility of fixing simultaneously  mean value and the variance
of the symmetry operator is a very important extension and ingredient of SCRPA. This has been worked
out for the first time in this present note. It implies that conservation laws and usual sum rules 
remain fulfilled in the same way as in the standard RPA. This is a very remarkable feature of SCRPA,
since the conservation of symmetries is a well known problem when attempting to include correlations 
beyond standard RPA. The SCRPA approach, contrary to the standard RPA which is perturbative, can be 
derived variationally and is thus non perturbative. We intend to apply the present theory to more
complicated situations and to give a more elaborate account of it in a future publication.\\     

{\bf Acknowledgments : }\\

Discussions with G. R\"opke are gratefully acknowledged.

\end{document}